\documentclass[onecolumn,preprintnumbers,pra,hyperref]{revtex4}
\usepackage{amssymb}
\usepackage{amsfonts}
\usepackage{amsmath}
\usepackage{graphicx}

\begin{document}

\title{Entangled state representation for deriving new operator identities regarding
to two-variable Hermite polynomial \thanks{{\small Work supported by the
National Natural Science Foundation of China under grant 10874174.}}}
\author{Hong-Yi Fan and Hong-Chun Yuan\thanks{Author to whom correspondence should be
addressed. Electronic mail: yuanhch@126.com or yuanhch@sjtu.edu.cn.}\\{\small Department of Physics, Shanghai Jiao Tong University, Shanghai
200240,} {\small \ China}}

\begin{abstract}
In this paper, by virtue of the entangled state representation we concisely
derive some new operator identities regarding to two-variable Hermite
polynomial (TVHP). By them and the technique of integration within an ordered
product (IWOP) of operators we further derive new generating function formulas
of TVHP. They are useful in quantum optical theoretical calculations. It is
seen from this work that by combining the IWOP technique and quantum
mechanical representations one can derive some new integration formulas even
without really performing the integration.


\end{abstract}

\maketitle

\section{Introduction}

It is well known that Hermite polynomial (HP) is an important special
function, which forms a class of orthogonal and complete set in function
space\cite{r1,r2,r3}. Among them, the two-variable Hermite polynomial (TVHP)
$H_{m,n}(\xi,\xi^{\ast})$\cite{r1,r4-1}
\begin{equation}
H_{m,n}(\xi,\xi^{\ast})=\sum_{l=0}^{\min[m,n]}\frac{(-1)^{l}m!n!\xi^{m-l}%
\xi^{\ast n-l}}{l!(m-l)!(n-l)!}\label{a1}%
\end{equation}
has become more and more useful (note that $H_{m,n}(\xi,\xi^{\ast})$
is not a direct product of two independent single-variable HP). For
examples, in quantum optics theory $H_{m,n}(\xi,\xi^{\ast})$ can be
considered as the basis of generalized Bargmann function space
corresponding to the two-mode Fock state\cite{b3}, and is very
useful for discussing quasi-probability distribution of some quantum
states\cite{b4}. In Fourier optics theory, $H_{m,n}(\xi,\xi^{\ast})$
is proved to be the eigenfunction of the complex fractional Fourier
transform, the corresponding phenomenon may be observed in the light
propagation in graded index (GRIN) medium\cite{b5}, this eigenmode
also exists in two-dimensional Talbot effect demonstrated in GRIN
medium\cite{b6}. In Ref.\cite{b1,b2}, it is pointed out that the
squeezed two-mode number state is just a TVHP excitation on the
two-mode squeezed vacuum state.

Recalling that the entangled state representation $\left \vert \xi \right \rangle
$ is expressed as\cite{r5}
\begin{equation}
\left \vert \xi \right \rangle =e^{-\frac{1}{2}\left \vert \xi \right \vert ^{2}+\xi
a^{\dagger}+\xi^{\ast}b^{\dagger}-a^{\dagger}b^{\dagger}}\left \vert
00\right \rangle ,\text{ }\xi=\xi_{1}+i\xi_{2},\label{a2}%
\end{equation}
which obeys%
\begin{equation}
\left(  a+b^{\dagger}\right)  \left \vert \xi \right \rangle =\xi \left \vert
\xi \right \rangle ,\text{ }\left(  a^{\dag}+b\right)  \left \vert \xi
\right \rangle =\xi^{\ast}\left \vert \xi \right \rangle ,\label{a3}%
\end{equation}
where $\left \vert 00\right \rangle $ is the two-mode vacuum state, $a^{\dag}$
and $b^{\dag}$ are the Bose creation operator with $\left[  a,a^{\dag}\right]
=\left[  b,b^{\dag}\right]  =1$. $\left \vert \xi \right \rangle $ is orthonormal
and complete,
\begin{equation}
\int \frac{d^{2}\xi}{\pi}\left \vert \xi \right \rangle \left \langle
\xi \right \vert =\int \frac{d^{2}\xi}{\pi}\colon e^{-\left(  a^{\dagger}%
+b-\xi^{\ast}\right)  \left(  a+b^{\dag}-\xi \right)  }\colon=1,\; \; \label{a4}%
\end{equation}
where $\colon \colon$ stands for normal ordering, and we have used the
technique of integration within an ordered product (IWOP) of
operators\cite{r7,r8,r9,r10,r11} as well as $\left \vert 00\right \rangle
\left \langle 00\right \vert =:\exp[-a^{\dagger}a-b^{\dag}b]:$ . According to
the generating function of $H_{m,n}\left(  \xi,\xi^{\ast}\right)  $%
\begin{equation}
\sum_{m,n=0}^{\infty}\frac{t^{m}t^{\prime n}}{m!n!}H_{m,n}\left(  \xi
,\xi^{\ast}\right)  =e^{-tt^{\prime}+t\xi+t^{\prime}\xi^{\ast}},\label{a5}%
\end{equation}
we can expand
\begin{equation}
\left \vert \xi \right \rangle =e^{-\frac{1}{2}\left \vert \xi \right \vert ^{2}%
}\sum_{m,n=0}^{\infty}\frac{a^{\dagger m}b^{\dagger n}}{m!n!}H_{m,n}\left(
\xi,\xi^{\ast}\right)  \left \vert 00\right \rangle ,\label{a6}%
\end{equation}
then the overlap between $\left \langle \xi \right \vert $ and the two-mode Fock
state is
\begin{equation}
\left \langle \xi \right \vert \left.  m,n\right \rangle =\frac{e^{-\frac{1}%
{2}\left \vert \xi \right \vert ^{2}}}{\sqrt{m!n!}}H_{m,n}^{\ast}\left(  \xi
,\xi^{\ast}\right)  ,\text{ }\left \vert m,n\right \rangle =\frac{a^{\dagger
m}b^{\dagger n}}{\sqrt{m!n!}}\left \vert 00\right \rangle .\label{a7}%
\end{equation}
Eq.(\ref{a6}) exhibits quantum entanglement\cite{r6}. In this work we shall
employ the entangled state representation for deriving some new operator
identities regarding to $H_{m,n}\left(  \xi,\xi^{\ast}\right)  $ and then
present their applications. By virtue of them, we can also derive some new
generating function formulas of $H_{m,n}\left(  \xi,\xi^{\ast}\right)  ,$
which is quite useful in calculating normalization of some quantum states. We
conclude that by combining the IWOP technique and quantum mechanical
representations one can derive some new integration formulas even without
really performing the integration.

\section{New operator identities regarding to two-variable Hermite
polynomials}

In order to obtain the normally ordered expansion of $H_{m,n}\left(
a+b^{\dagger},a^{\dag}+b\right)  ,$ by noticing $\left[  a+b^{\dagger}%
,a^{\dag}+b\right]  =0,$ we have
\begin{align}
e^{-tt^{\prime}+t\left(  a+b^{\dagger}\right)  +t^{\prime}\left(  a^{\dagger
}+b\right)  } &  =\colon e^{t\left(  a+b^{\dagger}\right)  +t^{\prime}\left(
a^{\dagger}+b\right)  }\colon \nonumber \\
&  =\sum_{m,n=0}^{\infty}\frac{t^{m}t^{\prime n}}{m!n!}\colon \left(
a+b^{\dagger}\right)  ^{m}\left(  a^{\dag}+b\right)  ^{n}\colon.\label{b1}%
\end{align}
where we have used Baker-Hausdorff formula%
\begin{equation}
e^{A}e^{B}=e^{B}e^{A}e^{\left[  A,B\right]  },\left[  A,\left[  A,B\right]
\right]  =\left[  B,\left[  A,B\right]  \right]  =0.
\end{equation}
Comparing with Eq.(\ref{a5}) we see the identity%
\begin{equation}
H_{m,n}\left(  a+b^{\dagger},a^{\dag}+b\right)  =\colon \left(  a+b^{\dagger
}\right)  ^{m}\left(  a^{\dag}+b\right)  ^{n}\colon.\label{b2}%
\end{equation}
Using Eqs.(\ref{a3}) and (\ref{a4}) as well as the IWOP
technique\cite{r7,r8,r9,r10,r11}, we have%
\begin{align}
H_{m,n}\left(  a+b^{\dagger},a^{\dag}+b\right)   &  =\int \frac{d^{2}\xi}{\pi
}H_{m,n}\left(  \xi,\xi^{\ast}\right)  \left \vert \xi \right \rangle
\left \langle \xi \right \vert \nonumber \\
&  =\int \frac{d^{2}\xi}{\pi}H_{m,n}\left(  \xi,\xi^{\ast}\right)  \colon
e^{-\left(  a^{\dagger}+b-\xi^{\ast}\right)  \left(  a+b^{\dag}-\xi \right)
}\colon \nonumber \\
&  =\colon \left(  a+b^{\dagger}\right)  ^{m}\left(  a^{\dag}+b\right)
^{n}\colon,\label{b3}%
\end{align}
which implies an integration formula%
\begin{equation}
\int \frac{d^{2}\xi}{\pi}H_{m,n}\left(  \xi,\xi^{\ast}\right)  \exp \left[
-\left(  \alpha^{\ast}-\xi^{\ast}\right)  \left(  \alpha-\xi \right)  \right]
=\alpha^{m}\alpha^{\ast n}.\label{b4}%
\end{equation}
Thus we see that by combining the IWOP technique and quantum mechanical
representations one can derive some new integration formulas even without
really performing the integration. Moreover, from the antinormally ordered
expansion (denoted by $\vdots$ $\vdots)$%
\begin{align}
e^{-tt^{\prime}+t^{\prime}\left(  a^{\dagger}+b\right)  +t\left(
a+b^{\dagger}\right)  } &  =\vdots e^{-2tt^{\prime}+t^{\prime}\left(
a^{\dagger}+b\right)  +t\left(  a+b^{\dagger}\right)  }\vdots \nonumber \\
&  =\sum_{m,n=0}^{\infty}\frac{2^{\left(  m+n\right)  /2}t^{m}t^{\prime n}%
}{m!n!}\vdots H_{m,n}\left(  \frac{a+b^{\dagger}}{\sqrt{2}},\frac{a^{\dag}%
+b}{\sqrt{2}}\right)  \vdots \label{b5}%
\end{align}
and comparing Eq.(\ref{a5}) with Eq.(\ref{b5}) we see%
\begin{equation}
H_{m,n}\left(  a+b^{\dagger},a^{\dag}+b\right)  =2^{\left(  m+n\right)
/2}\vdots H_{m,n}\left(  \frac{a+b^{\dagger}}{\sqrt{2}},\frac{a^{\dag}%
+b}{\sqrt{2}}\right)  \vdots.\label{11}%
\end{equation}
On the other hand, it is seen that
\begin{align}
&  \sum_{m,n=0}^{\infty}\frac{t^{m}t^{\prime n}}{m!n!}\left(  a+b^{\dagger
}\right)  ^{m}\left(  a^{\dag}+b\right)  ^{n}\nonumber \\
&  =e^{t\left(  a+b^{\dagger}\right)  }e^{t^{\prime}\left(  a^{\dagger
}+b\right)  }\nonumber \\
&  =\left.  \colon e^{tt^{\prime}+t^{\prime}\left(  b+a^{\dagger}\right)
}e^{t\left(  a+b^{\dagger}\right)  }\colon \right.  \nonumber \\
&  =\left.  \colon e^{-itit^{\prime}+t^{\prime}\left(  b+a^{\dagger}\right)
}e^{t\left(  a+b^{\dagger}\right)  }\colon \right.  \\
&  =\sum_{m,n=0}^{\infty}\frac{\left(  it\right)  ^{m}\left(  it^{\prime
}\right)  ^{n}}{m!n!}\colon H_{m,n}\left[  -i\left(  a+b^{\dagger}\right)
,-i\left(  a^{\dag}+b\right)  \right]  \colon,\label{b6}%
\end{align}
this leads to the new operator identity%
\begin{equation}
\left(  a+b^{\dagger}\right)  ^{m}\left(  a^{\dag}+b\right)  ^{n}%
=i^{m+n}\colon H_{m,n}\left[  -i\left(  a+b^{\dagger}\right)  ,-i\left(
a^{\dag}+b\right)  \right]  \colon.\label{b8}%
\end{equation}
Similarly, using Eqs.(\ref{a3}) and (\ref{a4}), we also have%
\begin{align}
\left(  a+b^{\dagger}\right)  ^{m}\left(  a^{\dag}+b\right)  ^{n} &
=\int \frac{d^{2}\xi}{\pi}\xi^{m}\xi^{\ast n}\left \vert \xi \right \rangle
\left \langle \xi \right \vert \nonumber \\
&  =\int \frac{d^{2}\xi}{\pi}\xi^{m}\xi^{\ast n}\colon \exp \left[  -\left(
a^{\dagger}+b-\xi^{\ast}\right)  \left(  a+b^{\dag}-\xi \right)  \right]
\colon \nonumber \\
&  =i^{m+n}\colon H_{m,n}\left[  -i\left(  a+b^{\dagger}\right)  ,-i\left(
a^{\dag}+b\right)  \right]  \colon,\label{b7}%
\end{align}
which implies another integration formula%
\begin{equation}
\int \frac{d^{2}\xi}{\pi}\xi^{m}\xi^{\ast n}\exp \left[  -\left(  \alpha^{\ast
}-\xi^{\ast}\right)  \left(  \alpha-\xi \right)  \right]  =i^{m+n}%
H_{m,n}\left(  -i\alpha,-i\alpha^{\ast}\right)  .\label{b9}%
\end{equation}
This is the reciprocal relation of Eq.(\ref{b4}). It is clear shown that Eqs.
(\ref{b4}) and (\ref{b9}) are the mutual integration transformation between
$H_{m,n}\left(  \xi,\xi^{\ast}\right)  $ and the polynomial $\xi^{m}\xi^{\ast
n}.$

To derive another more complicated generating function formula about TVHP$,$
we introduce another two-mode entangled state involved in modes $c^{\dag
},d^{\dagger}$,
\begin{equation}
\left \vert \zeta \right \rangle =e^{-\frac{1}{2}\left \vert \zeta \right \vert
^{2}+\zeta c^{\dag}+\zeta^{\ast}d^{\dagger}-c^{\dag}d^{\dagger}}\left \vert
00\right \rangle ,\text{ }\zeta=\zeta_{1}+\text{i}\zeta_{2},\label{b10}%
\end{equation}
where $\left[  c,c^{\dag}\right]  =1$ and $\left[  d,d^{\dagger}\right]  =1,$
obeying%
\begin{equation}
\left(  c+d^{\dagger}\right)  \left \vert \zeta \right \rangle =\zeta \left \vert
\zeta \right \rangle ,\text{ }\left(  c^{\dag}+d\right)  \left \vert
\zeta \right \rangle =\zeta^{\ast}\left \vert \zeta \right \rangle ,\label{b11}%
\end{equation}
and
\begin{equation}
\int \frac{d^{2}\zeta}{\pi}\left \vert \zeta \right \rangle \left \langle
\zeta \right \vert =\int \frac{d^{2}\zeta}{\pi}\colon e^{-\left(  c^{\dagger
}+d-\zeta^{\ast}\right)  \left(  c+d^{\dag}-\zeta \right)  }\colon
=1.\; \; \label{b12}%
\end{equation}
As a result of Eq.(\ref{b3}) we deduce%
\begin{align}
G &  \equiv \sum_{m,n=0}^{\infty}\frac{t^{n}s^{m}}{n!m!}H_{m,n}\left(
a+b^{\dagger},a^{\dag}+b\right)  H_{m,n}\left(  c+d^{\dagger},c^{\dag
}+d\right)  \nonumber \\
&  =\sum_{m,n=0}^{\infty}\frac{t^{n}s^{m}}{n!m!}\colon \left(  a+b^{\dagger
}\right)  ^{m}\left(  a^{\dag}+b\right)  ^{n}\left(  c+d^{\dagger}\right)
^{m}\left(  c^{\dag}+d\right)  ^{n}\colon \nonumber \\
&  =\colon \exp \left[  s\left(  a+b^{\dagger}\right)  \left(  c+d^{\dagger
}\right)  +t\left(  a^{\dag}+b\right)  \left(  c^{\dag}+d\right)  \right]
\colon.\label{b13}%
\end{align}
In reference to
\begin{equation}
\int \frac{d^{2}z}{\pi}e^{\eta \left \vert z\right \vert ^{2}+fz+gz\ast}=-\frac
{1}{\eta}e^{-fg/\eta},\text{ }\operatorname{Re}\eta<0,\label{b14}%
\end{equation}
the right-hand side of Eq.(\ref{b13}) can be the result of the following
integration with use of the IWOP technique%
\begin{align}
G &  =\int \frac{d^{2}\zeta}{\pi}\colon e^{-|\zeta|^{2}+\zeta \left[  \left(
c^{\dag}+d\right)  +s\left(  a+b^{\dagger}\right)  \right]  +\zeta^{\ast
}\left[  \left(  c+d^{\dagger}\right)  +t\left(  a^{\dag}+b\right)  \right]
}\nonumber \\
&  \times e^{-st\left(  a^{\dag}+b\right)  \left(  a+b^{\dagger}\right)
-\left(  c^{\dag}+d\right)  \left(  c+d^{\dagger}\right)  }\colon \nonumber \\
&  =\colon \int \frac{d^{2}\zeta}{\pi}\left \vert \zeta \right \rangle \left \langle
\zeta \right \vert e^{-st\left(  a^{\dag}+b\right)  \left(  a+b^{\dagger
}\right)  +s\zeta \left(  a+b^{\dagger}\right)  +t\zeta^{\ast}\left(  a^{\dag
}+b\right)  }\colon \label{b15}%
\end{align}
where, within the normal ordering $:$ $:$, the terms involving $\left(
a+b^{\dagger}\right)  $ and $\left(  a^{\dag}+b\right)  $ can be expressed as
the following integration%
\begin{align}
&  e^{-st\left(  a^{\dag}+b\right)  \left(  a+b^{\dagger}\right)
+s\zeta \left(  a+b^{\dagger}\right)  +t\zeta^{\ast}\left(  a^{\dag}+b\right)
}\nonumber \\
&  =\frac{1}{1-ts}\int \frac{d^{2}\xi}{\pi}\exp \left[  \frac{-|\xi|^{2}}%
{1-ts}+\xi \left(  a^{\dag}+b+\frac{s\zeta}{1-ts}\right)  \right.  \nonumber \\
&  \left.  +\xi^{\ast}\left(  a+b^{\dagger}+\frac{t\zeta^{\ast}}{1-ts}\right)
-\frac{ts|\zeta|^{2}}{1-ts}-\left(  a^{\dag}+b\right)  \left(  a+b^{\dagger
}\right)  \right]  .\label{b16}%
\end{align}
Substituting Eq.(\ref{b16}) into Eq.(\ref{b15}) and using Eqs.(\ref{a3}) and
(\ref{a4}) as well as Eqs.(\ref{b11}) and (\ref{b12}) we see%
\begin{align}
G &  =\frac{1}{1-ts}\int \frac{d^{2}\xi}{\pi}\int \frac{d^{2}\zeta}{\pi
}\left \vert \xi \right \rangle \otimes \left \vert \zeta \right \rangle \left \langle
\zeta \right \vert \otimes \left \langle \xi \right \vert \nonumber \\
&  \times \exp \left[  \frac{-ts\left(  |\xi|^{2}+|\zeta|^{2}\right)  }%
{1-ts}+\frac{s\zeta \xi}{1-ts}+\frac{t\zeta^{\ast}\xi^{\ast}}{1-ts}\right]
\nonumber \\
&  =\frac{1}{1-ts}\exp \left \{  \frac{-ts}{1-ts}\left[  \left(  a^{\dag
}+b\right)  \left(  a+b^{\dagger}\right)  +\left(  c^{\dag}+d\right)  \left(
c+d^{\dagger}\right)  \right]  \right.  \nonumber \\
&  \left.  +\frac{s\left(  a+b^{\dagger}\right)  \left(  c+d^{\dagger}\right)
}{1-ts}+\frac{t\left(  c^{\dag}+d\right)  \left(  a^{\dag}+b\right)  }%
{1-ts}\right \}  .\label{b17}%
\end{align}
Since $\left(  a^{\dag}+b\right)  ,\left(  a+b^{\dagger}\right)  ,\left(
c^{\dag}+d\right)  $ and $\left(  c+d^{\dagger}\right)  $ are all commutative
among themselves, we can make replacement $\left(  a^{\dag}+b\right)
\rightarrow y,\left(  a+b^{\dagger}\right)  \rightarrow x,\left(  c^{\dag
}+d\right)  \rightarrow$ $y^{\prime}$, $\left(  c+d^{\dagger}\right)
\rightarrow x^{\prime}$, then by comparing Eq.(\ref{b17}) with Eq.(\ref{b13})
we obtain the complicated generating function formula about TVHP
\begin{equation}
\sum_{m,n=0}^{\infty}\frac{s^{m}t^{n}}{m!n!}H_{m,n}\left(  x,y\right)
H_{m,n}\left(  x^{\prime},y^{\prime}\right)  =\frac{1}{1-ts}\exp \left[
\frac{sxx^{\prime}+tyy^{\prime}-ts\left(  xy+x^{\prime}y^{\prime}\right)
}{1-ts}\right]  .\label{b18}%
\end{equation}

\section{Applications}

We now present some applications of Eq.(\ref{b18}). To begin with, we point
out that Eq.(\ref{b18}) can be used for deriving another generating function
formula. In fact,\ using the generating function of Laguerre
polynomial\cite{r3}%
\begin{equation}
\sum_{m=0}^{\infty}L_{m}\left(  x\right)  s^{m}=\left(  1-s\right)  ^{-1}%
\exp \left(  \frac{-xs}{1-s}\right)  ,\label{d1}%
\end{equation}
and its relation to TVHP%
\begin{equation}
H_{m,m}\left(  x,y\right)  =\left(  -1\right)  ^{m}m!L_{m}\left(  xy\right)
,\label{d2}%
\end{equation}
we can reexpress Eq.(\ref{b18}) as%
\begin{align}
&  \sum_{m,n=0}^{\infty}\frac{s^{m}t^{n}}{m!n!}H_{m,n}\left(  x,y\right)
H_{m,n}\left(  x^{\prime},y^{\prime}\right)  \nonumber \\
&  =\frac{e^{tyy^{\prime}}}{1-st}\exp \left[  \frac{-\left(  xy+x^{\prime
}y^{\prime}-tyy^{\prime}-\frac{1}{t}xx^{\prime}\right)  ts}{1-st}\right]
\nonumber \\
&  =e^{tyy^{\prime}}\sum_{m=0}^{\infty}\left(  st\right)  ^{m}L_{m}\left(
xy+x^{\prime}y^{\prime}-tyy^{\prime}-\frac{1}{t}xx^{\prime}\right)
\nonumber \\
&  =e^{tyy^{\prime}}\sum_{m=0}^{\infty}\frac{\left(  -st\right)  ^{m}}%
{m!}H_{m,m}\left[  i(\sqrt{t}y^{\prime}-\frac{x}{\sqrt{t}}),i(\sqrt{t}%
y-\frac{x^{\prime}}{\sqrt{t}})\right]  .\label{d3}%
\end{align}
Comparing the same power of $s$ on the above two sides yields%
\begin{equation}
\sum_{n=0}^{\infty}\frac{t^{n}}{n!}H_{m,n}\left(  x,y\right)  H_{m,n}\left(
x^{\prime},y^{\prime}\right)  =\left(  -t\right)  ^{m}e^{tyy^{\prime}}%
H_{m,m}\left[  i(\sqrt{t}y^{\prime}-\frac{x}{\sqrt{t}}),i(\sqrt{t}%
y-\frac{x^{\prime}}{\sqrt{t}})\right]  .\label{d5}%
\end{equation}
Further, by noticing%
\begin{align}
e^{t^{\prime}a}e^{ta^{\dagger}} &  =e^{tt^{\prime}}e^{ta^{\dagger}%
}e^{t^{\prime}a}=\colon e^{\left(  -\text{i}t^{\prime}\right)  \text{i}%
a+\left(  -\text{i}t\right)  \text{i}a^{\dagger}-\left(  -\text{i}t\right)
\left(  -\text{i}t^{\prime}\right)  }\colon \nonumber \\
&  =\sum_{m,n=0}^{\infty}\frac{\left(  -\text{i}t\right)  ^{m}\left(
-\text{i}t^{\prime}\right)  ^{n}}{m!n!}\colon H_{m,n}\left(  \text{i}a^{\dag
},\text{i}a\right)  \colon \label{c1}%
\end{align}
and comparing it with
\begin{equation}
e^{t^{\prime}a}e^{ta^{\dagger}}=\sum_{m,n=0}^{\infty}\frac{t^{\prime n}t^{m}%
}{n!m!}a^{n}a^{\dagger m},\label{c2}%
\end{equation}
we have the compact operator identity%
\begin{equation}
a^{n}a^{\dag m}=\left(  -\text{i}\right)  ^{m+n}\colon H_{m,n}\left(
\text{i}a^{\dag},\text{i}a\right)  \colon.\label{c3}%
\end{equation}
It then follows from Eq.(\ref{b18}) that%
\begin{align}
e^{sab}e^{ta^{\dag}b^{\dagger}} &  =\sum_{m=0}^{\infty}\frac{s^{m}}{m!}%
a^{m}b^{m}\sum_{n=0}^{\infty}\frac{t^{n}}{n!}a^{\dag n}b^{\dag n}\nonumber \\
&  =\sum_{m,n=0}^{\infty}\frac{\left(  -1\right)  ^{m+n}s^{m}t^{n}}%
{m!n!}\colon H_{m,n}\left(  \text{i}a^{\dag},\text{i}a\right)  H_{m,n}\left(
\text{i}b^{\dag},\text{i}b\right)  \colon \nonumber \\
&  =\frac{1}{1-ts}\colon \exp \left[  \frac{ts\left(  a^{\dag}a+b^{\dagger
}b\right)  +sa^{\dag}b^{\dagger}+tab}{1-ts}\right]  \colon \nonumber \\
&  =\frac{1}{1-ts}e^{\frac{s}{1-ts}a^{\dag}b^{\dagger}}e^{-\left(  a^{\dag
}a+b^{\dagger}b\right)  \ln \left(  1-ts\right)  }e^{\frac{t}{1-ts}%
ab}\label{c4}%
\end{align}
where at the last step we have used\cite{r10}
\begin{equation}
e^{\lambda a^{\dag}a}=\colon \exp[\left(  e^{\lambda}-1\right)  a^{\dag
}a]\colon.\label{c5}%
\end{equation}
This result of Eq.(\ref{c4}) agree with that of Ref.\cite{r12}. On the other
hand, using the antinormally ordered operator $\vdots H_{m,n}\left(  a^{\dag
},a\right)  \vdots$ $=a^{\dag m}a^{n}$, where using its P-representation as
well as Eq.(\ref{b4}) we easily prove that
\begin{equation}
\vdots H_{m,n}\left(  a^{\dag},a\right)  \vdots=\int \frac{d^{2}z}{\pi}%
H_{m,n}\left(  z^{\ast},z\right)  \left \vert z\right \rangle \left \langle
z\right \vert =\int \frac{d^{2}z}{\pi}H_{m,n}\left(  z^{\ast},z\right)  \colon
e^{-\left(  a^{\dagger}-z^{\ast}\right)  \left(  a-z\right)  }\colon=a^{\dag
m}a^{n},
\end{equation}
we can obtain%
\begin{align}
e^{ta^{\dag}b^{\dagger}}e^{sab} &  =\sum_{m,n=0}^{\infty}\frac{t^{n}s^{m}%
}{n!m!}a^{\dag n}a^{m}b^{\dagger n}b^{m}\nonumber \\
&  =\sum_{m,n=0}^{\infty}\frac{s^{m}t^{n}}{m!n!}\vdots H_{m,n}\left(
a,a^{\dag}\right)  H_{m,n}\left(  b,b^{\dag}\right)  \vdots \nonumber \\
&  =\frac{1}{1-ts}\vdots \exp \left[  \frac{-ts\left(  a^{\dag}a+b^{\dagger
}b\right)  +ta^{\dag}b^{\dagger}+sab}{1-ts}\right]  \vdots \nonumber \\
&  =\frac{1}{1-ts}e^{\frac{s}{1-ts}ab}\vdots e^{\frac{-ts}{1-ts}\left(
a^{\dag}a+b^{\dagger}b\right)  }\vdots e^{\frac{t}{1-ts}a^{\dag}b^{\dagger}%
}.\label{c6}%
\end{align}
Finally, according to Eqs.(\ref{d2}), (\ref{d5}) and (\ref{c3}), we also have%
\begin{align}
a^{m}b^{m}e^{a^{\dag}b^{\dagger}\tanh \lambda} &  =\sum_{n=0}^{\infty}%
\frac{\tanh^{n}\lambda}{n!}a^{m}a^{\dag n}b^{m}b^{\dag n}\nonumber \\
&  =\left(  -1\right)  ^{m}\sum_{n=0}^{\infty}\frac{\left(  -\tanh
\lambda \right)  ^{n}}{n!}\colon H_{m,n}\left(  \text{i}a,\text{i}a^{\dag
}\right)  H_{m,n}\left(  \text{i}b,\text{i}b^{\dag}\right)  \colon \nonumber \\
&  =m!\tanh^{m}\lambda e^{a^{\dag}b^{\dag}\tanh \lambda}\colon L_{m}\left(
-a^{\dag}a-b^{\dag}b-ab\coth \lambda-a^{\dag}b^{\dag}\tanh \lambda \right)
\colon,\label{d4}%
\end{align}
which relates to Laguerre polynomial. Based on this, we obtain that the
two-mode photon-subtracted squeezed vacuum state is expressed as\cite{r14}
\begin{align}
\left \vert \lambda \right \rangle _{m} &  \equiv a^{m}b^{m}S_{2}(\lambda
)\left \vert 00\right \rangle \nonumber \\
&  =a^{m}b^{m}e^{a^{\dag}b^{\dagger}\tanh \lambda}\left \vert 00\right \rangle
\nonumber \\
&  =m!\tanh^{m}\lambda e^{a^{\dag}b^{\dagger}\tanh \lambda}L_{m}\left(
-a^{\dag}b^{\dag}\tanh \lambda \right)  \left \vert 00\right \rangle \nonumber \\
&  =m!\tanh^{m}\lambda L_{m}\left(  -a^{\dag}b^{\dag}\tanh \lambda \right)
S_{2}(\lambda)\left \vert 00\right \rangle ,\label{d6}%
\end{align}
where $S_{2}(\lambda)=\exp \left[  \lambda(a^{\dag}b^{\dagger}-ab)\right]  $ is
the two-mode squeezing operator with $\lambda$ being a real squeezing
parameter. From Eq.(\ref{d6}), $\left \vert \lambda \right \rangle _{m}$ can is
equivalent to Laguerre polynomial excitation on squeezed vacuum state. Recall
in Ref.\cite{r15}, we have calculated its normalization factor%
\begin{equation}
\left \langle 00\right \vert e^{ab\tanh \lambda}a^{\dag m}b^{\dag m}a^{m}%
b^{m}e^{a^{\dag}b^{\dagger}\tanh \lambda}\left \vert 00\right \rangle =\left(
m!\right)  ^{2}\sinh^{2m}\lambda P_{m}\left(  \cosh2\lambda \right)  \label{d7}%
\end{equation}
where $P_{m}\left(  x\right)  $ is Legendre polynomial\cite{r1,r14}%
\begin{equation}
P_{m}(x)=\sum_{l=0}^{\left[  m/2\right]  }\frac{\left(  -1\right)  ^{l}\left(
2m-2l\right)  !x^{m-2l}}{2^{m}l!(m-l)!(m-2l)!}.\label{d8}%
\end{equation}
Using Eqs.(\ref{d6}) and (\ref{d7}) as well as the coherent state's
completeness relation $\int \frac{d^{2}\alpha d^{2}\beta}{\pi^{2}}\left \vert
\alpha,\beta \right \rangle \left \langle \alpha,\beta \right \vert =1,$ it follows%
\begin{equation}
\int \frac{d^{2}\alpha d^{2}\beta}{\pi^{2}}L_{m}\left(  -\alpha \beta
\tanh \lambda \right)  L_{m}\left(  -\alpha^{\ast}\beta^{\ast}\tanh
\lambda \right)  e^{-\left \vert \alpha \right \vert ^{2}-\left \vert
\beta \right \vert ^{2}+\left(  \alpha \beta+\alpha^{\ast}\beta^{\ast}\right)
\tanh \lambda}=\cosh^{2m}\lambda P_{m}\left(  \cosh2\lambda \right)  \label{d9}%
\end{equation}
which is a new integration formula.

In summary, by virtue of the entangled state representation we concisely
derive some new operator identities regarding to two-variable Hermite
polynomials. They are useful in quantum optical theoretical calculations. By
combining the IWOP technique and quantum mechanical representations one can
derive some new integration formulas even without really performing the integration.

\end{document}